\documentclass[%
 aip,
 jmp,%
 amsmath,amssymb,
 reprint,%
 nofootinbib,
floatfix,
]{revtex4-1}

\usepackage{graphicx} 
\usepackage{mhchem}
\usepackage{multirow}
\usepackage{array}
\usepackage{float}
\usepackage{physics}
\usepackage{amsfonts, amsmath}
\usepackage{amssymb,bm,bbm}
\usepackage{xcolor}
\usepackage{hyperref}

\newcommand{\eqnref}[1]{Eq.~\eqref{#1}}
\newcommand{\figref}[1]{Fig.~\ref{#1}}
\newcommand{\secref}[1]{Sec.~\ref{#1}}
\newcommand{\appref}[1]{Appendix~\ref{#1}}

\newcommand{\citeref}[1]{Ref.~\citenum{#1}}

\newcommand{\cminv}{cm${}^{-1}$}
\newcommand{\etal}{\nobreakspace\emph{et al.}}

\usepackage{caption}  

\date{January 2025}

\begin{document}
\title{Infrared spectroscopy of protonated water clusters via the quantum thermal bath method and highly accurate machine-learned potentials}
\author{T. Baird}
 \affiliation{Centre Europ\'een de Calcul Atomique et Mol\'eculaire (CECAM), Ecole Polytechnique F\'ed\'erale de Lausanne, 1015 Lausanne, Switzerland}
\author{R. Vuilleumier}%
\affiliation{ 
PASTEUR, D\'epartement de chimie, \'Ecole normale sup\'erieure, PSL University, Sorbonne Universit\'e. CNRS, 75005 Paris, France
}%
\author{S. Bonella}
\email{sara.bonella@epfl.ch}
 \affiliation{Centre Europ\'een de Calcul Atomique et Mol\'eculaire (CECAM), Ecole Polytechnique F\'ed\'erale de Lausanne, 1015 Lausanne, Switzerland}
\date{\today}

\begin{abstract}

The spectral features of water clusters provide important information on their structure and dynamics and can assist in deciphering the nature of the local environment of aqueous solutions in a variety of different conditions. Accurately capturing these features via numerical simulations is a non-trivial task that typically requires a sophisticated combination of high-level electronic structure methods and costly quantum dynamics techniques. We present results of molecular dynamics simulations of the IR spectra of protonated water clusters, ranging from the monomer to the tetramer, obtained via the combination of highly accurate machine-learned potential energy surfaces (PES) and dipole moment surfaces (DMS), and  the quantum thermal bath (QTB) methodology which facilitates cost-effective inclusion of NQEs in molecular dynamics simulations. We compare our results with previous theoretical and experimental studies and show that this combination provides a significantly cheaper, yet still suitably accurate, alternative to more traditional computational approaches.

\end{abstract}
\maketitle

\section{Introduction}

The investigation of water clusters holds significant theoretical and practical importance due to the ubiquitous presence of water in nature and its crucial role in many chemical and biological processes. The  spectral properties of water clusters offer insights not only into the structure and dynamics of these clusters but carry also signatures that help in understanding the behavior of bulk water across different environments. Furthermore, systematic studies of small water clusters also provide information about proton transfer, a process of key importance in chemistry and physics. Finally, the infrared (IR) spectra of water clusters are highly sensitive to their internal hydrogen bonding network, making IR spectroscopy a valuable tool for probing intermolecular interactions amongst the water molecules which constitute these clusters \cite{li2020infrared,ohno2005effect}.

Accurately modelling these intermolecular interactions is essential for capturing the various molecular motions that give rise to spectral features. As a result, in computational studies, high-precision potential energy surfaces and force fields are required for reliable spectral calculations. Similarly, the accuracy of dipole moment estimates is critical for faithfully reproducing the spectral peak positions and relative intensities. This necessity, in turn, demands sophisticated and computationally expensive electronic structure methods to accurately determine the system’s charge density and dipole moments. The high computational cost associated with \textit{first principles} calculations has led to considerable efforts to develop sophisticated analytical models for the potential and dipole moment surfaces for different systems~\cite{Wang2011, Cisneros2016, Reddy2017, yu2017high, Yu2017a,Schwan2019}. More recently, machine learning techniques have emerged as a powerful alternative to  design many body potentials from accurate electronic structure calculations, in particular for water from clusters to bulk\cite{morawietz2012neural,morawietz2013density,natarajan2015representing,jindal2025computing,morawietz2016van,zhang2021phase,maxson2024transferable,montero2024comparing,omranpour2024perspective,Bore2023}.

Incorporating nuclear quantum effects (NQEs) is also essential for accurately reproducing key infrared spectral features of water clusters \cite{yu2019classical}. These include the characteristic red shift of peaks relative to classical predictions and the presence of combination bands. Different numerical strategies have been adopted to account for NQEs. Notably, spectra have been computed using the multi-configurational time-dependent Hartree (MCTDH) method to solve the nuclear Schrödinger equation \cite{schroder2022coupling,vendrell2007full,pelaez2017infrared}. These simulations, however, do not account for temperature effects and alternative approaches based either on semiclassical methods~\cite{liu2009quantum,benson2021matsubara} or path integral molecular dynamics (PIMD)~\cite{trenins2019path, rossi2014remove} have been proposed. While these approaches do provide interesting insights, they are known to face limitations due the combination of classical dynamics and quantum probability distribution sampling (semiclassical) or may become quite expensive in deep quantum regimes and fail to reproduce, for example, the correct trends for peaks' intensities at low temperature without bespoke corrections~\cite{ benson2021matsubara}.

In this paper, we revisit the spectroscopy of protonated water clusters at finite temperature combining a state-of-the-art ML representation of the interactions and the dipole moment surface recently developed in the group of D. Marx~\cite{schran2019automated,beckmann2022infrared} with the quantum thermal bath (QTB) \cite{qtb} approach. We compare the resulting spectra with those obtained from classical MD simulations and with previous numerical and experimental studies to evaluate the accuracy of this approach and assess the influence of NQEs on the spectral properties of water clusters.

\section{Methods}
\label{sec:theory_and_methods}

\subsection{Quantum thermal bath dynamics}
QTB models quantum effects by coupling nuclear classical dynamics to a thermal reservoir endowed with a noise power spectral density that emulates quantum fluctuations. In practice, this involves solving numerically the generalized Langevin dynamics : \begin{align} \label{eq:qtb_eom} \begin{split} &m_I \dv[2]{\vb{R}_I}{t} = - \nabla_{\vb{R}_I} E({\vb{R}_I}) - m_I \gamma \dv{\vb{R}_I}{t} + \boldsymbol{F}_I(t) \end{split} \end{align} In the equation above, ${\vb{R}_I}$ are the Cartesian coordinates of ion $I$, and $m_I$ is the mass of the ion. $\gamma$ is the friction coefficient of the thermostat, and $\boldsymbol{F}_I = [F_I^{x}\ F_I^y \ F_I^z]$ represents a three-dimensional random force acting on ion $I$, with the associated colored noise spectrum: \begin{equation} \label{eq:qtb_fdt} C _{F^{\alpha}_J F^{\beta}_K} (\omega) = 2m_K \gamma \theta(\omega, T) \delta _{JK} \delta ^{\alpha \beta} \qquad \alpha, \beta \in {x,y,z } \end{equation} where $C _{F^{\alpha}_J F^{\beta}_K}(\omega)$ denotes the autocorrelation function between the $\alpha$-component of the force on ion $J$ and the $\beta$-component on ion $K$. The function \begin{equation} \label{eq:qtb_spectrum} \theta(\omega, T) = \frac{\hbar \omega}{2\tanh(\frac{\hbar \omega}{2 k_B T})} \end{equation} is the quantum mechanical power spectral density of a harmonic oscillator at temperature $T$. The QTB evolution, which is exact only for harmonic systems, can be justified based on an approximation to the path integral representation of the density matrix of a generic quantum system linearly coupled to a bath of quantum harmonic oscillators\cite{schmid1982quasiclassical}. Within the limits of this justification, QTB has proven accurate in a number of applications\cite{bronstein2014quantum, mauger2021nuclear,qtb}. Most notably, the approach performs well for the simulation of vibrational spectroscopy\cite{ple2021anharmonic}. A well-documented limitation of the Quantum Thermal Bath (QTB), also shared by other semiclassical approaches, is its susceptibility to zero-point energy leakage (ZPEL) \cite{brieuc2016zero}. This issue arises when the energy residing in high-frequency vibrational modes—initially distributed according to the quantum thermal statistics encoded by QTB— cascades into lower-frequency modes due to the classical time evolution that drives towards equipartition. An effective approach proposed to alleviate ZPEL is the adaptive quantum thermal bath (adQTB)~\cite{mangaud2019fluctuation}. adQTB monitors ZPEL on the fly via the diagnostic function proposed in~\cite{mangaud2019fluctuation}: 
\begin{equation} 
\label{eq:qtb_delta_fdt}
 \Delta_{\text{FDT},i}(\omega) = \text{Re}[C_{vF,i}(\omega)] - m_i\gamma_i(\omega) C_{vv,i}(\omega)
 \end{equation}
 where $C_{vF,i}(\omega)$ and $C_{vv,i}(\omega)$ denote the cross-correlation between velocity and random force, and the velocity autocorrelation, respectively, for the $i$-th Cartesian coordinate. Deviations from zero of the function above indicate violations of the quantum fluctuation dissipation theorem induced by ZPEL. These can then be mitigated by dynamically adjusting the friction coefficient $\gamma$ in a frequency-dependent fashion, i.e., $\gamma = \gamma(\omega)$, with the goal of minimizing the magnitude of $\Delta_{\text{FDT},i}(\omega)$ across all modes and degrees of freedom~\cite{mangaud2019fluctuation}. 
 
 Details on the QTB numerical implementation adopted in this work are presented in Appendix~\ref{sec:Appendix_QTB_Implementation}. Notably, calculations presented in the next section did not show appreciable ZPEL, with the fluctuation-dissipation relation (\eqnref{eq:qtb_delta_fdt}) proving to be well-satisfied, and the IR spectra obtained using QTB and adQTB being in very good agreement for all systems considered. Consequently, in the following we report only the results obtained with standard QTB.

\subsection{NN-PES/DMS and computation of IR spectra}
To ensure a fast and accurate representation of the interactions and of the dipole surface for the water clusters, we adopted specifically designed machine-learned force fields and dipoles. We incorporated the Neural Network Potential (NNP) library \cite{schran2019automated,beckmann2022infrared} (built on RubNNet4MD \cite{Brieuc2020}) into the Tinker-HP molecular dynamics package \cite{lagardere2018tinker}. The Tinker-HP engine had a pre-existing extensive suite of functionality for both classical and quantum MD simulations, in particular providing support for QTB and adaptive QTB (adQTB), limiting our task to interfacing the NNP library.

The NNP library is based on the neural network architecture introduced by Behler and Parrinello \cite{behler2007generalized} and has been trained on a comprehensive dataset of water cluster configurations (ranging from \ce{H2O} to \ce{H9O4+}) generated using ab initio molecular dynamics (AIMD) and path integral AIMD. The energies of these clusters were computed at a level of accuracy comparable to coupled-cluster CCSD(T) theory. The NNP library yields accurate and transferable energies and forces for these systems, and it has been employed in multiple studies to successfully predict the static and dynamical properties of water clusters~\cite{schran2019automated}. Additionally, the library includes a dipole moment surface (DMS) trained on a dataset of molecular configurations, with dipole moments computed at coupled-cluster accuracy. Use of the neural network drastically reduces the substantial computational cost associated with directly performing these high-level quantum chemistry calculations
. The NNP-DMS approach thus enables an efficient yet accurate determination of charge distributions and dipole moments, which are essential for infrared (IR) spectral calculations \cite{beckmann2022infrared, larsson2022state}.

In all calculations reported in the following, the IR spectrum is obtained from the (Kubo-transformed) dipole derivative autocorrelation function. Specifically, during each simulation step, the total dipole moment of the system is predicted by passing the ionic geometry through the DMS. The dipole time derivative is then computed using finite differences and the Fourier transform of the dipole derivative autocorrelation function, $C _{\dot{\mathbf{\mu}}\dot{\mathbf{\mu}}}(\omega)$, is subsequently determined.
For QTB simulations, this autocorrelation function is Kubo-transformed using:

\begin{equation} \tilde{C} _{\dot{\mathbf{\mu}}\dot{\mathbf{\mu}}}(\omega) = \frac{\tanh(\beta \hbar \omega/2)}{\hbar \omega} C _{\dot{\mathbf{\mu}}\dot{\mathbf{\mu}}}(\omega) \end{equation}
and the IR spectrum is then computed using the Beer-Lambert law:

\begin{equation} \label{eq:beer_lambert} n(\omega) \alpha(\omega) = \frac{\pi \beta}{3 c V \varepsilon_0} \tilde{C} _{\dot{\mathbf{\mu}}\dot{\mathbf{\mu}}}(\omega) \end{equation}
where $n(\omega)$ is the refractive index, $\alpha(\omega)$ is the absorption coefficient, $c$ is the speed of light, $V$ is the simulation box volume, and $\varepsilon_0$ is the vacuum permittivity.

In this work, the final IR spectra are obtained by averaging the spectra derived from trajectory segments covering a frequency range of 0-15,000 \cminv{} (spanning far-infrared to near-infrared regions, including overtone ranges). Each segment is smoothed using a Hann window function of width 750 \cminv{} in order to mitigate ringing artefacts, and the initial trajectory segments (corresponding to approximately 25 ps of simulation time) are discarded to account for equilibration. Lastly, unless otherwise stated, the final calculated spectrum is itself smoothed via a sliding window procedure, with the window size set to 50 \cminv{}. This window size was chosen such that it is sufficiently small to avoid loss of resolution of any fine spectral features whilst simultaneously reducing the noisiness of the spectra.

Since theoretical IR spectra in the literature are often computed with the center-of-mass (COM) motion removed, we implemented functionality to eliminate COM motion in our simulations for direct comparison with existing studies. In our classical MD simulations, this was done conventionally by subtracting translational and rotational contributions of the COM motion from atomic velocities at each timestep. For QTB, however, this procedure is more delicate due to the need to preserve the quantum fluctuation-dissipation relation, which relies on all degrees of freedom. Details on our algorithm are provided in \appref{sec:com_removal}.

\section{RESULTS} 

We have performed simulations of several systems of increasing size, namely: \ce{H2O}, \ce{H3O+}, \ce{H(H2O)_2+}, \ce{H(H2O)_3+}, \ce{H(H2O)_4+} (Eigen), \ce{H(H2O)_4+} (ring, \textit{cis-}, and \textit{trans}-Zundel). Computed IR spectra of a selection of these systems (\ce{H2O}, \ce{H(H2O)_2+}, \ce{H(H2O)_3+} and \ce{H(H2O)_4+} (Eigen)) are presented in the following sections, together with analysis of the main spectral features observed. The remaining clusters are analyzed in \appref{app:water_clusters_ir_extras}. We also report the compute time required for the calculation of IR spectra for the full set of 8 clusters in \appref{sec:watercluster_timings}.

In all cases, simulations were performed in the gas phase and within the NVT ensemble, with different temperatures considered to allow comparison with existing experimental and theoretical results. The simulations were performed for a total of 1 ns (discarding the first 25 ps as equilibration), the timestep was set to dt=0.1 fs, and properties recorded every 10 fs. The small value of the timestep was found to improve the resolution of the spectral features in the higher frequency ranges considered. The ability to affordably obtain such long trajectory lengths facilitates significant improvements in spectral resolution and statistics.  In the classical simulations, the temperature was controlled via Langevin dynamics, integrated using the BAOAB thermostat\cite{leimkuhler2015molecular}, with a friction coefficient of 2 ps${}^{-1}$. This choice of friction allowed for a good balance between enhancing ergodicity whilst minimizing the damping effect on the system's dynamics, thus maintaining resolution of the IR spectrum. In all plots, the spectra are normalized so that the highest peak of the figure takes on the value 1.

\label{sec:waterclusters_results}
\subsection{Water monomer (\ce{H2O})}

To verify that we have correctly coupled the NNP library with the engine for performing quantum and classical MD simulations, and to carry out a preliminary evaluation the accuracy of this combination of methodologies, we first performed simulations of a single neutral water molecule in the gas phase. We take as reference results the reported theoretical IR spectra of Benson \etal{} \cite{benson2020quantum} and Ple \etal{} \cite{ple2021anharmonic}, both of whom computed the infrared spectrum of a single gas phase water molecule at 300K using a diversity of approximate and exact methodologies for quantum dynamics and sampling. In particular, they report spectra calculated via: classical MD, thermostatted ring polymer MD (TRPMD) \cite{rossi2014remove}, centroid molecular dynamics (CMD) \cite{witt2009applicability}, quasi-centroid molecular dynamics (QCMD) \cite{trenins2019path}, various linearized semiclassical initial value representation (LSC-IVR) approaches \cite{liu2009quantum}, and the discrete variable representation (DVR) \cite{tennyson2004dvr3d}. They take the latter as the exact reference result. Moreover, Ple \etal{} also report the spectra calculated via the use of QTB and adQTB simulations of the water monomer, which we will use as a check for our own results, although we note that the dipole moment surface and forcefield used in our simulations are different to those employed in their study.

In order to affect a comparison with the reference results, we do not remove the motion of the center-of-mass of the system (neither translational nor rotational). As a consequence, we observe that all peaks within the calculated spectra gain an additional sub-peak which originates from the combination with global rotations.

\begin{figure*}[ht]
  \centering
  \includegraphics[width=1.0\textwidth]{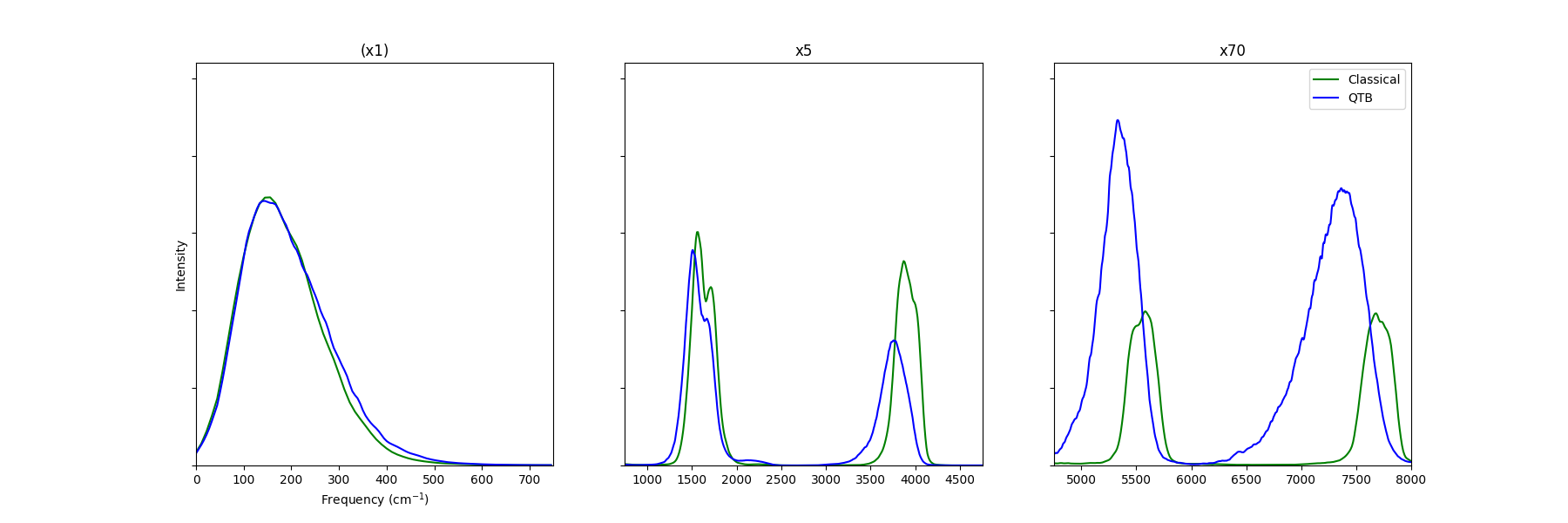}
  \caption[IR spectrum of a single water molecule in the gas phase at 300K.]
  {IR spectrum of a single water molecule in the gas phase at 300K. Classical results are shown in green while the spectrum obtained via QTB is shown in blue. Scaling of the y-axis is the same as in~\cite{benson2020quantum}.}
  \label{fig:water_IR}
\end{figure*}

The results of the classical and QTB simulations are presented in Fig.\ref{fig:water_IR}. From this figure, we observe that our QTB results are for the most part in very good agreement with the exact reference and previous theoretical results of \citeref{benson2020quantum} (see Fig.1 in reference) and \citeref{ple2021anharmonic} (see Fig.4 of reference), with the positions and relative intensities of the peaks in the IR spectrum being accurately predicted.

Firstly, considering the frequency range 0-4000 \cminv{}, we observe that QTB correctly captures the red-shift of the peak at $\sim$3700 \cminv{} (corresponding to water stretches) with respect to the classical prediction. This is a well-known consequence of incorporating NQEs into calculations of IR spectra. Moreover, the position of the peak is found to be in good agreement with the exact (DVR) reference. This peak is somewhat broadened in the QTB approaches and the symmetric and antisymmetric contributions are not perfectly resolved, much like what is observed in the TRPMD and LSC-IVR results of \citeref{benson2020quantum}.

We also note the presence of the bend-libration combination band around 2000 \cminv{} in the QTB spectrum, which is a result of the coupling of the low frequency librational modes with the OH bending mode, located at approximately 1600 \cminv{}. This band is not present in the classical spectrum, and its presence in the QTB spectrum is a result of the inclusion of NQEs in the calculations together with the accuracy of the dipole moment surface used, highlighting the importance of these factors in the accurate prediction of the IR spectrum of water.

Furthermore, we observe that the relative magnitudes of the various peaks in the region $\omega < 4500$ \cminv{} are also in good agreement with the DVR results. 

Turning to the near-infrared region ($\omega>$  4700 \cminv{}), there are two combination bands at $\sim$5300 \cminv{} and $\sim$7300 \cminv{} which we again observe are correctly downshifted in frequency in the QTB spectra compared to the classical spectra. Benson \etal{} had previously remarked that the intensities of these peaks are severely underestimated by ring polymer methods, with their intensities no more accurate than the classical results, which they find to drastically undershoot the relative intensities seen in the exact spectrum. Moreover, the authors found that the results obtained via the LSC-IVR method gave considerably better agreement of the relative intensities with the exact results. Based on these observations, they concluded that the failure of ring polymer methods in the near infrared region should not be due to a lack of real-time coherences, as LSC-IVR also neglects these. 

In the case of QTB, we in fact find that our predicted relative intensities are in very close agreement with the exact reference, overestimating only very slightly both the 5300 \cminv{} and 7300 \cminv{} peaks, compared to the bending and stretching bands. This observation is aligned with that made in a separate study which employed (ad-)QTB to compute the IR spectrum of gas phase water \cite{ple2021anharmonic}. In this work, the authors proposed that the explanation for the greater accordance of the (ad-)QTB results with the exact reference in the near-infrared region is due to the fact that QTB provides a more faithful sampling of the underlying Wigner distribution than ring polymer methods, highlighting the importance of appropriately accounting for correlations in the nuclear phase space distribution as well as coherences in the quantum dynamics. \\Note that, although still underestimated, our classical predictions of the relative intensities are noticeably closer to the exact results in this region than those reported in \cite{benson2020quantum}. A possible explanation for this disparity is the different dipole moment surfaces and/or potential energy surfaces used in the two studies. \\Overall, the agreement of our QTB results with the exact reference spectra for \ce{H2O} across the entirety of the frequency range considered is very good, with both peak locations and widths being accurately predicted.

\subsection{Protonated water dimer/Zundel cation (\ce{H(H2O)_2+})}
\label{sec:zundel_ir}

As the next step-up in complexity, simulations of the gas-phase protonated water dimer \ce{H(H2O)_2+} were performed at a temperature of 300K. To assess the accuracy of our simulations, we compare with the results of \citeref{huang2008comparison} by Huang \textit{et al}. In this work, the authors performed a series of simulations of the Zundel cation using MCTDH, classical MD, RPMD, normal mode analysis (NMA), and the MULTIMODE (MM) method \cite{carter2000vibrations}. At variance with the simulations of the monomer, we have removed the rotational and translational motion of the center-of-mass (COM) of the system in order to match as closely as possible the conditions of this reference work.

\begin{figure*}[ht]
  \centering
  \includegraphics[width=1.0\textwidth]{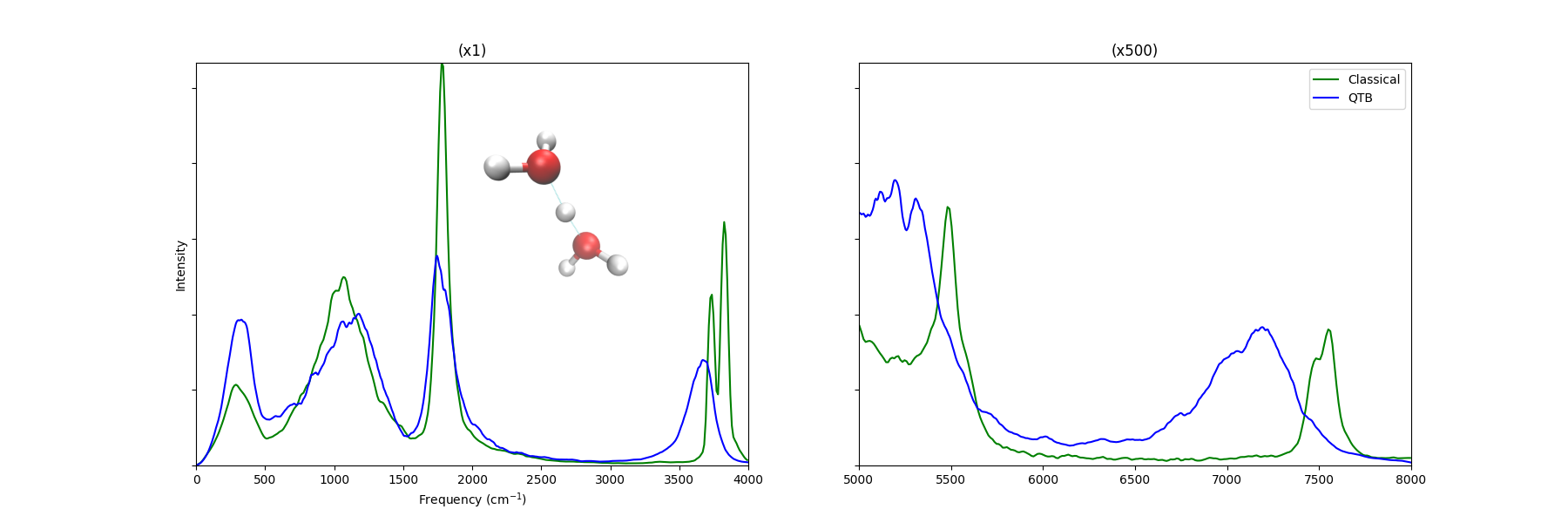}

  \caption[IR spectrum of the protonated water dimer in the gas phase at 300K.]{IR spectrum of the protonated water dimer in the gas phase at 300K. Our results are scaled with respect to the maximum peak in our computed spectra so that the y-axis is in arbitrary units and has a maximum value of 1.Classical results are shown in green while the spectrum obtained via QTB is shown in blue.} 
  \label{fig:zundel_300k}
\end{figure*}

To assess the effectiveness of our approach, we consider each of the main spectral features in the frequency range 0-4000 \cminv{} as shown in the left-hand side panel of \figref{fig:zundel_300k}, before briefly turning our attention to the overtone region 5000-8000 \cminv{} illustrated in the right-most panel of \figref{fig:zundel_300k}.

The first observation that can be made is with regard to the peak at around 1100-1200 \cminv{}, which, as remarked by Huang \etal{}, has previously been attributed to an asymmetric proton stretch mode. This mode is found to be considerably blue-shifted in reference quantum results with respect to harmonic results. Our classical MD result for this mode is in good agreement with that of Huang \etal{}, with a peak around 1100 \cminv{}, and the QTB spectra shows a blue-shift similar to the RPMD result in this reference work, yielding a peak at around 1200 \cminv{}. Similarly to the RPMD result of Huang \etal{}, the QTB spectrum lacks the doublet-like structure observed in the MCTDH spectra at around 900-1000 \cminv{} (see lower panel of Fig.2 in \citeref{huang2008comparison}). Notably, zero-temperature MCTDH results \cite{vendrell2007full} show no intensity in the region around 1200 \cminv{}, with a doublet around 1000 \cminv{} and peaks around 1400 \cminv{} and 1900 \cminv{}. 

The blue shift of the proton-transfer peak can be explained by the nearly quartic form of the potential energy along the asymmetric proton stretch\cite{mouhat_thermal_2023}. The classical dynamics probes only the bottom of the well where the harmonic frequency is low, while the quantum dynamics also probes the repulsive walls. As for the comparison with MCTDH results, it is interesting to note that averaging the MCTDH peaks' positions with the spectral weights reported in \cite{vendrell2007full}, we find that the resulting weighted average peak position is in good agreement with that which we observe in our QTB spectrum, and which was previsouly also observed in RPMD spectra \cite{huang2008comparison}. The broad QTB feature could then originate from thermal effects which lead to a coalescing of the distinct peaks observed in the MCTDH spectrum. We moreover observe that the relative intensities of the peaks (both in relation to one another and between methodologies) are in fairly good accordance with those reported by Huang \etal{} The lowest lying peak, slightly below 500 \cminv{}, (corresponding to wagging, rocking and twisting modes) is slightly more intense in the QTB spectrum relative to the classical compared to the relative intensities observed between the classical and RPMD reference results, whereas the opposite is true for the peak at around 1100-1200 \cminv{}.

\begin{figure}[H]
  \centering
  \includegraphics[width=0.5\textwidth]{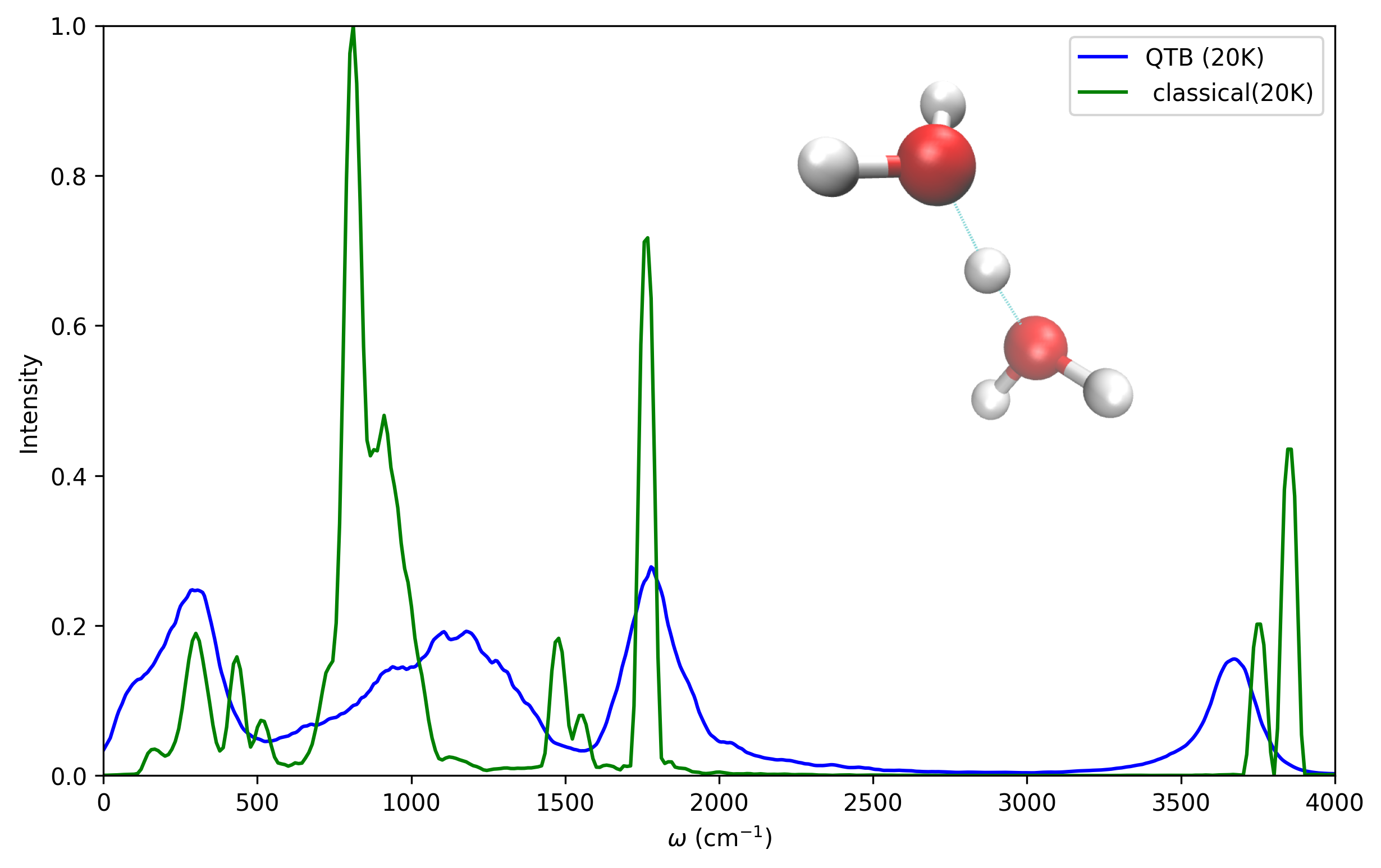}
  \caption[IR spectrum of the protonated water dimer in the gas phase at 20K.]{IR spectrum of the protonated water dimer in the gas phase at 20K. This lower temperature was considered in order to investigate whether the doublet structure observed in the MCTDH spectrum of \ce{H(H2O)_2+} at around 1700 \cminv{} would emerge in the classical and QTB spectra when thermal effects are reduced. Classical results are shown in green while the spectrum obtained via QTB is shown in blue.}
  \label{fig:zundel_20k}
\end{figure}

The peaks observed at $\sim$1700 \cminv{} (attributable to water bending \cite{vendrell2007dynamics}) in our classical and QTB results are in good agreement with the classical MD and RPMD results of Huang \etal{} respectively. The peak in the QTB spectrum is very slightly red-shifted with respect to the classical MD result, in accordance with the RPMD reference. Moreover, the MCTDH spectrum of \ce{H(H2O)_2+}, reported in a separate work by Vendrell \etal{} \cite{vendrell2007dynamics}, indicates a doublet structure at this frequency which again is not observed in the QTB results, nor the classical or RPMD results of Huang \etal{} Again this may be attributable to broadening induced by thermal effects and NQEs. Indeed, if we lower the temperature to 20K, we observe the emergence of this doublet feature in our classical spectra (together with a number of less prominent, yet distinct, peaks - see \figref{fig:zundel_20k}). The broadening of the peaks in the QTB spectra persists even to these low temperatures, however, and is a consequence the zero-point motion, captured by QTB, and which endows the system with a significant ``effective temperature'' even at the low physical temperature of 20K.

Finally, turning our attention to the frequency region corresponding to O-H stretches (approximately 3000-4000 \cminv{}), we observe that the peak in the QTB spectrum at around 3700 \cminv{} is red-shifted with respect to the classical (and harmonic) prediction which assumes a doublet structure with peaks located at $\sim$3730 \cminv{} and $\sim$3825 \cminv{}. This is in accordance with the behaviour of the RPMD result in the reference work, which exhibits a similar red-shift. However, at variance with the RPMD and classical results (alongside the MCTDH and harmonic calculations), which exhibit a doublet-like structure in this region, the QTB result shows a single broad peak. This is again likely due to the broadening effect that the QTB thermostat has on the spectrum.

Before concluding our analysis of the spectrum of the protonated water dimer, we turn our attention to the overtone region of the IR spectrum, which is presented in the right-hand side panel of \figref{fig:zundel_300k}. Firstly, we observe that both our classical MD and QTB results correctly capture the overtone peaks at around 7500 \cminv{} (7200 \cminv{}) and 5500 \cminv{} (5300 \cminv{}) which have been previously reported in an experimental work by McDonald \etal{} \cite{mcdonald2018near}, and whose locations can also be predicted on theoretical grounds based on the positions of lower-lying fundamentals. The first of these overtones is attributed to twice the frequency of the O-H stretch mode, whilst the second is attributed to the combination of the O-H stretch and water bending modes. The combination of the O-H stretch and asymmetric proton stretch modes are less clear (being only vaguely discernible in the classical spectrum and less so in the QTB spectrum), likely due to the fact this overtone is somewhat subsumed into the broad combination band of the O-H stretch and water bending modes. We see that the doublet structure of the classical MD result for the O-H stretch is preserved in its overtone. Moreover, QTB correctly predicts the red-shift of these overtone peaks with respect to the classical MD results.

To summarise,  QTB is able to accurately predict the main spectral features of the Zundel cation. Peak locations and relative intensities are in good accordance with reference RPMD results within the frequency range considered. QTB fails to capture the doublet structure in the O-H stretch region which is ostensibly a consequence of the broadening effect of the thermostat causing these sub-peaks to coalesce, but otherwise we find that even the higher frequency spectral features found in the overtone region of the spectrum are in good agreement with their predicted locations.

\subsection{Protonated water trimer (\ce{H(H2O)_3+})}
\label{sec:h7o3}

For this cluster, we primarily report results for the lower temperature of 100K in order to compare with the thorough spectroscopic analysis of the protonated water trimer performed by Yu \etal{} \cite{yu2019classical}. In this study, the authors performed a series of RPMD, classical MD, double harmonic, and quantum VSCF/VCI \cite{roy2013vibrational} calculations with the goal of comparing the observed spectral features of \ce{H(H2O)_3+} with previous experimental results. For the sake of completeness, afterwards we also present the results of our simulations at 300K in Fig.\ref{fig:h7o3_300K}. In order to achieve satisfactory temperature control a friction value of 20 ps${}^{-1}$ was used in the classical MD simulations and to counteract the broadening effect this has on the spectra, the deconvolution procedure first introduced by Rossi \etal{} was employed \cite{rossi2018fine,supp_mauger2021nuclear}.

\begin{figure}[H]
  \centering
    \includegraphics[width=0.5\textwidth]{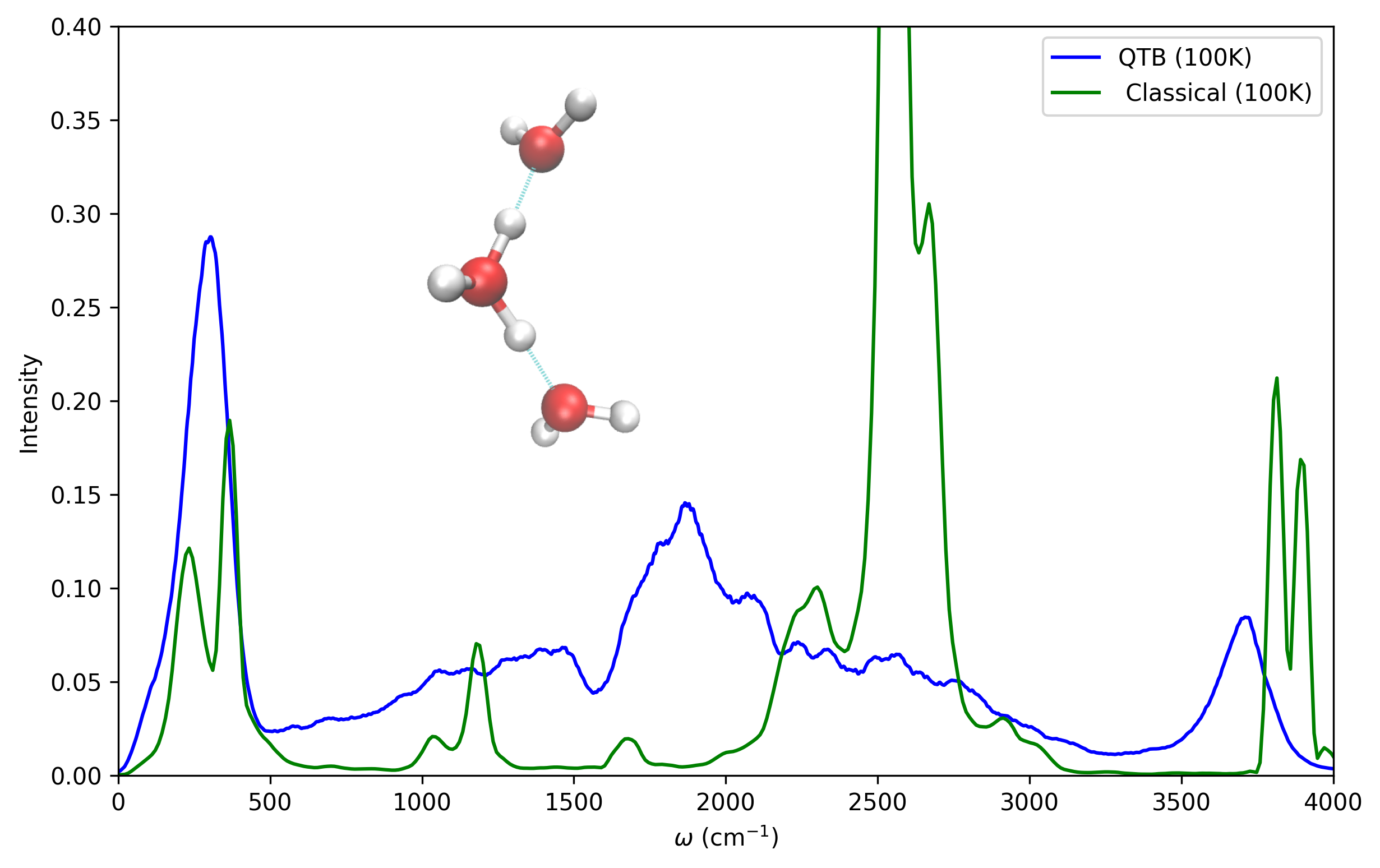}
    \caption[IR spectrum of the protonated water trimer in the gas phase.]{IR spectrum of the protonated water trimer in the gas phase. All simulations were performed at 100K. Classical results are shown in green while the spectrum obtained via QTB is shown in blue.}
    \label{fig:h7o3_IR}
\end{figure}

Despite considering a lower temperature of 20K via classical MD, the reference work only reports RPMD for the higher temperature of 100K. This is due to the considerable computational cost associated with the necessity of employing a larger number of nuclear replicas in the RPMD simulations at lower temperatures to reach convergence. Here we therefore have a nice example of the utility of QTB, in that it facilitates the inclusion of NQEs even at very low temperatures, where the cost of RPMD proves to be a deterrent to its use. Indeed, we were able to perform QTB simulations of the trimer at the reduced temperature of 20K without issue. However, given the negligible differences in the resulting spectrum compared to that obtained at 100K, we opt to show only the latter.

Let us first consider the main (double) peak at around 2600 \cminv{} in the classical MD spectrum presented in Fig.\ref{fig:h7o3_IR}. This band is known to result from stretching modes associated with the proton attached to the hydrated hydronium core of the cluster, with the lower frequency peak corresponding to the asymmetric stretch and the higher frequency peak corresponding to the symmetric stretch mode \cite{duong2017disentangling}. Our classical results at 100K are in good agreement with the reference classical IR spectrum at the same temperature, with the location of the peaks corresponding closely to the double harmonic results owing to the low temperature. Our classical MD results, however, present considerably more pronounced peaks in comparison with the reference. This is likely due to our use of deconvolution for the QTB results that, for the sake of consistency, was applied also to our classical results. QTB also reproduces this peak, which is again considerably broadened. The peak is red-shifted with respect to the classical MD and harmonic predictions, in agreement with the RPMD results of Yu \etal{} \cite{yu2019classical}. This down-shift is in fact closer to that observed in the VSCF/VCI and experimental reference results (see Fig.2(e-f) in \cite{yu2019classical}), moving the peak closer to the experimental value of $\sim$1900 \cminv{}. This is a notable success of combining QTB with this highly-accurate NN-PES and NN-DMS. Given that the classical MD results we obtain are in good accordance with those of Yu \etal{} this would suggest that it is the use of QTB which is more likely to be responsible for this superior agreement than it is the particulars of the PES and DMS. 

We note that, as is also the case with TRPMD\cite{yu2019classical}, the QTB spectrum is significantly broadened in the frequency range 1500-2500 \cminv{} and therefore does not reproduce the fine-grained spectral features present in the experimental results and which are well-captured by VSCF/VCI. Neither use of deconvolution nor reduction of the friction coefficient in concert with adQTB (results not shown) was sufficient to resolve these fine-grained features.
Lastly, we note that the two features observed at 240 \cminv{} and 350 \cminv{} in the reference work appear to have merged in the QTB results, most probably due to broadening induced by the thermostat.

Spectra of \ce{H(H2O)_3+} calculated at the elevated temperature of 300K are presented in Fig.\ref{fig:h7o3_300K}. These are essentially broadened versions of the spectra at 100K, with the broadening most prominent in the classical case.
The right-hand side panel of Fig.\ref{fig:h7o3_300K} presents a zoom-in on the overtone region of the IR spectrum of the protonated water trimer. We observe overtones at approximately 5300 \cminv{} and 7300 \cminv{} in the QTB spectra and 5500 \cminv{} and 7600 \cminv{} in the classical MD spectrum. The positions of these bands are as expected, being the sum of the hydronium stretch band ($\sim$1900 \cminv{}) and the water stretch band ($\sim$3700 \cminv{}) and twice the frequency of the water stretch mode respectively, with the typical red-shift of the quantum results being correctly reproduced by QTB.

\begin{figure*}[ht]
    \centering
    \includegraphics[width=1.0\textwidth]{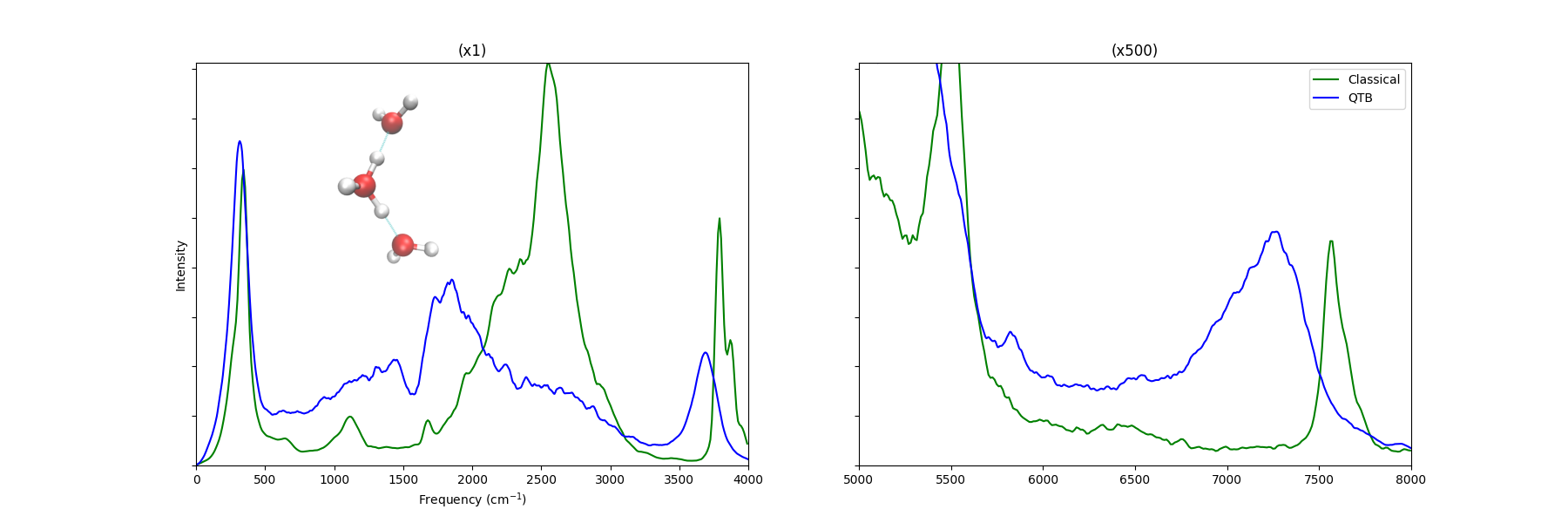}
    \caption{IR spectrum of the protonated water trimer in the gas phase at 300K.Classical results are shown in green while the spectrum obtained via QTB is shown in blue.}
    \label{fig:h7o3_300K}
\end{figure*}

To consolidate these observations, we find that QTB does a good job of reproducing the major spectral features of the protonated water trimer. The peak locations predicted by QTB for the most prominent bands in this cluster's IR spectrum are in generally good agreement with the experimental reference and prior theoretical results using highly sophisticated and computationally expensive methodologies. For one of the principal bands, namely the doublet at $\sim$2600 \cminv{} arising from proton stretching modes, we even find that QTB apparently provides a better estimate of the red-shift of this band relative to the classical reference than has previously been found with RPMD. 
Similarly to observations made for the dimer system, we once more find that many of the less intense and more fine-grained features which are present in reference VSCF/VCI results are missing in the QTB spectrum due to thermal broadening and amalgamation of these bands in the latter.

\subsection{Protonated water tetramer (\ce{H(H2O)_4+}) in the Eigen conformation.}
\label{sec:eigen_spectra}

\begin{figure}[H]
  \centering
    \includegraphics[width=0.5\textwidth]{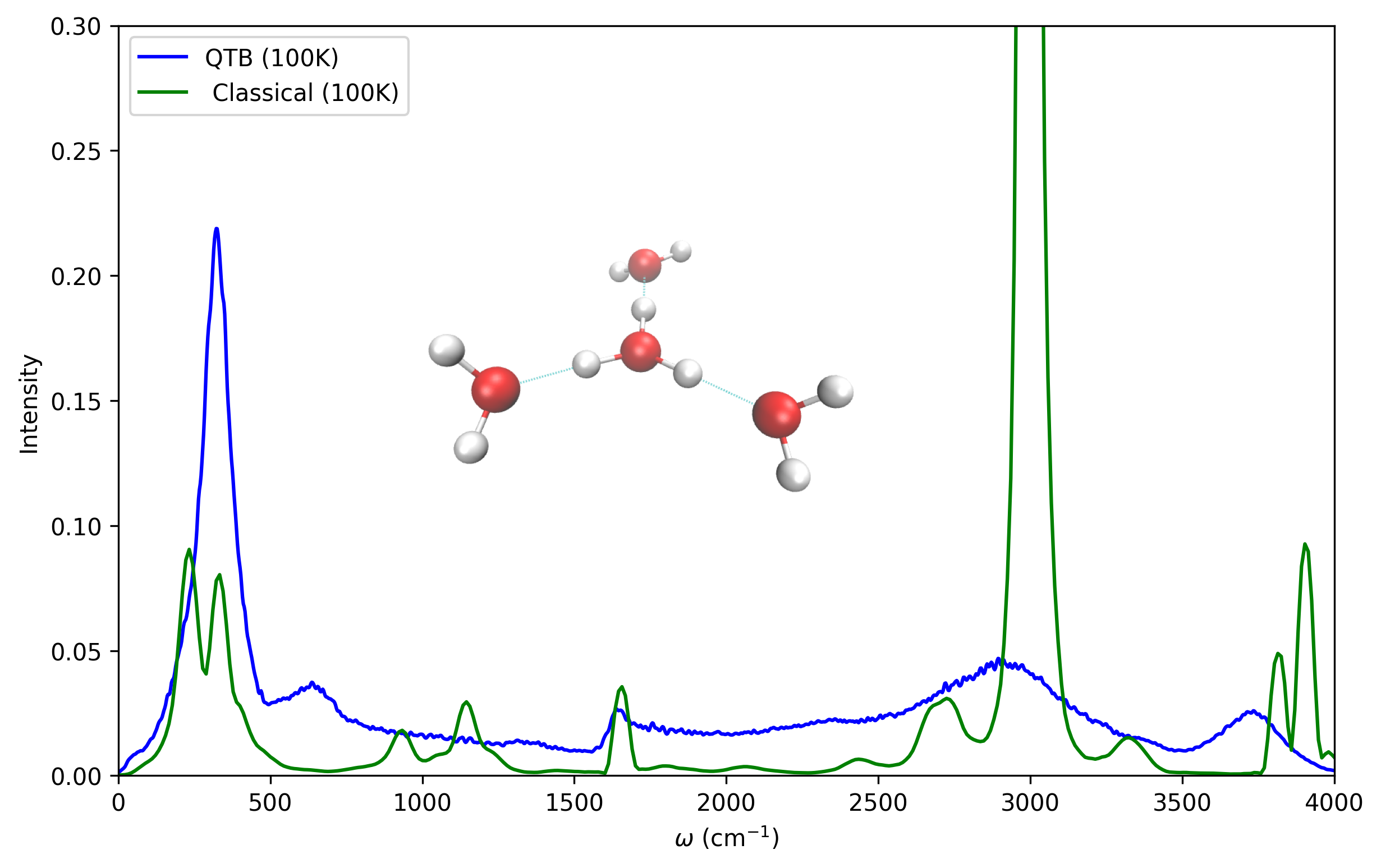}

    \caption[IR spectra of the protonated water tetramer in the gas phase.]{IR spectrum of the protonated water tetramer in the gas phase at 100K. Classical results are shown in green while the spectrum obtained via QTB is shown in blue.}
    \label{fig:h9o4_IR_20k}
\end{figure}

To further test the capabilities of our calculations, we simulated of the gas phase protonated water tetramer in the Eigen conformation at three different temperatures: 20K, 100K, and 300K. The lower temperature simulations were carried out in order to enable comparison with reference experimental and theoretical works. The theoretical results are again taken from a series of papers by Yu \etal{} \cite{yu2019classical, yu2017high} which employed RPMD, classical MD, double harmonic, and VSCF/VCI methods to compute the IR spectra of the protonated Eigen isomer of \ce{H(H2O)_4+}, whilst the experimental results (which are also used by Yu \etal{} as a reference) were taken from the works of Esser \etal{} \cite{esser2018deconstructing} and Wolke \etal{} \cite{wolke2016spectroscopic}. As in the case of the protonated water trimer, the reference TRPMD results were limited to the higher temperature of 100K due to the issue of high computational expense incurred by RPMD at lower temperatures. On the other hand, the reference experimental results were obtained at $T\approx 20K$. As we are primarily interested in the comparison of the QTB results with the reference TRPMD results, we present only the results of our QTB and classical MD simulations at 100K. The calculated spectra are shown in \figref{fig:h9o4_IR_20k}. As in the case of the water trimer, we find that the spectra at 20K are not appreciably different from those at 100K, and so we do not present them here.

The spectrum of the protonated water tetramer is very reminiscent of that of the protonated water trimer which has just been discussed.
The most prominent feature in the IR spectrum of the tetramer is the peak at around 2650 \cminv{} (in the experimental spectrum), which has been attributed to the asymmetric stretch of the hydronium core \cite{yu2019classical}. 
As in the TRPMD results reported by Yu \etal{} \cite{yu2019classical}, although the QTB results do demonstrate a red-shift of this peak with respect to the classical MD and double harmonic results, this shift is not sufficiently large to match the experimental location of the peak at around 2650 \cminv. Rather, QTB predicts a peak at approximately 2880 \cminv{}, a non-negligible overestimation of this mode's frequency.

On the other hand, we note that the band corresponding to water stretches at around 3750 \cminv{} is in good agreement with the TRPMD results of Yu \etal{}, exhibiting a red-shift with respect to the classical MD and double harmonic results by approximately 200 \cminv{}. At variance with the experimental result, this band is not split into two distinct peaks in the QTB result, and is instead broadened considerably. This is similar to what the authors of the cited work observe in the case of TRPMD. Similarly, all the lower intensity peaks lying between 1500-2500 \cminv{}, which are present in the experimental spectrum, and which are picked up by VSCF/VCI, are not discernible in the QTB spectrum due to broadening and merging of these bands, again like the behaviour observed in the case of TRPMD. Likewise, the far-infrared region of the spectrum is not well-resolved by QTB, nor classical MD, with only a single broad peak being observed between 0-500 \cminv{}. This is in contrast to the reference experimental and QCMD results (see Fig 7.(e) and Fig.7 (g-h) in \cite{yu2019classical}), which exhibit a number of distinct peaks in this region. These peaks have been attributed to the torsional and stretching motion of the lateral water molecules present in \ce{H(H2O)_4+}  \cite{esser2018deconstructing}.

\begin{figure*}[ht]
    \centering
    \includegraphics[width=1.0\textwidth]{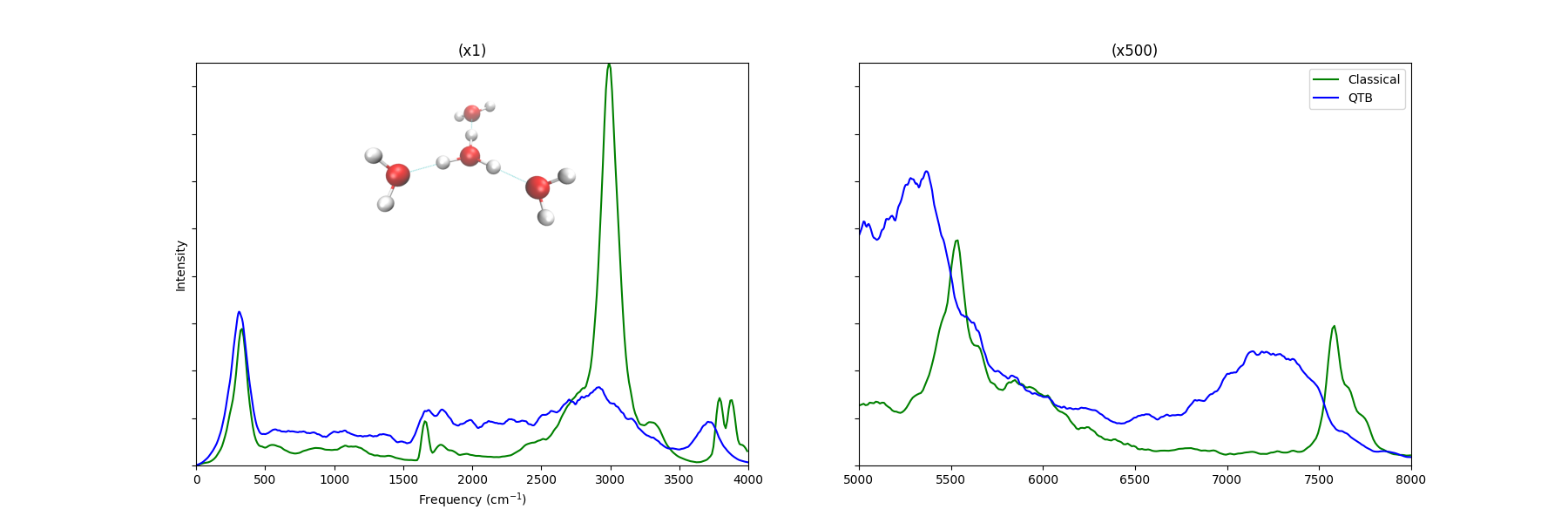}
    \caption{IR spectrum of the protonated water tetramer in the gas phase at 300K. Classical results are shown in green while the spectrum obtained via QTB is shown in blue.}
    \label{fig:h9o4_300K}
\end{figure*}

Finally, for the sake of completeness, we present the IR spectra of the tetramer at 300K in \figref{fig:h9o4_300K} with the overtone region shown in the rightmost panel. As in the case of the protonated water trimer, these are essentially broadened versions of the spectra at 100K, with the broadening most prominent in the classical case given the aforementioned presence of zero-point motion offsetting the reduction in temperature when nuclear quantum effects are included. The overtones are found to be located at their expected frequencies, with QTB again correctly predicting the red-shift of these peaks with respect to the classical MD results.

To summarize our analysis of the Eigen cation, we find that QTB also performs well in capturing the main spectral characteristics of this larger water cluster. The predicted positions of the major bands in the IR spectrum show good overall agreement with both experimental data and previous high-level theoretical studies. As in the previous case of the trimer, the QTB spectrum lacks many of the finer, low-intensity features seen in VSCF/VCI reference results, with these features instead seeming to contribute to broad bands in the frequency ranges in which they are observed in the reference spectra. Regardless, comparison with the classical reference once again indicates the success of QTB in reproducing the characteristic red-shifts of the main vibrational modes when accounting for nuclear quantum effects, with these shifts also clearly visible in the overtone region of the spectrum.

\section{Conclusions}

In this work, we have combined cost-effective QTB dynamics with accurate and efficient machine-learned force fields and dipole moment surfaces to compute the infrared spectra of protonated water clusters. The spectral features of clusters of  increasing size and complexity were obtained and compared with available theoretical and experimental results. Our IR spectra were found to be in generally good agreement with references. The locations and relative intensities of the main spectral features of the clusters were accurately computed for all clusters considered. The inclusion of NQEs by means of QTB was shown to be essential in order to capture typical red-shifts of the peaks with respect to the classical prediction. Incorporation of NQEs with QTB also allows one to obtain a good description of various combination bands and overtones, both in position and intensities. The main limitation of QTB is that it misses some of the fine structure of the spectra due to broadening induced by coupling to the bath. This can be mitigated with deconvolution, but not always to a sufficient extent to enable complete resolution of the fine spectral structure. Crucially, the computational cost of QTB calculations is essentially identical to that of a classical simulation and considerably less than the computational cost of traditional ring polymer MD and alternative methods. The results presented in this paper demonstrate the potential of combining QTB with a high accuracy machine-learned force field and dipole moment surface and pave the way for future calculations with similar set-ups.

\acknowledgments
We are grateful to Christoph Schran and Dominik Marx for help with the potential energy surface (PES) and the dipole moment surface (DMS) at CCSD(T) accuracy used in this work, which are both represented in the framework of high-dimensional neural networks. The authors also thank Simon Huppert and Thomas Pl\'e for useful discussions on the use of QTB for IR spectroscopy and assistance with the Tinker HP code.
This research was supported by the NCCR MARVEL, a National Centre for Competence in Research, funded by the Swiss National Science Foundation (grant number 205602).

\bibliography{refs}

\appendix
\section{Implementing QTB dynamics}
\label{sec:Appendix_QTB_Implementation}

\eqnref{eq:qtb_eom}, \eqnref{eq:qtb_fdt}, and \eqnref{eq:qtb_spectrum} define a generalized Langevin dynamics framework, effectively modeling the interaction with an Ohmic bath --- i.e., a bath characterized by a spectral density that scales linearly with frequency. In practical implementations, the stochastic force must be prepared in advance of integrating the equation of motion \eqref{eq:qtb_eom}. The noise is typically generated on-the-fly (OTF), following the method introduced in Ref.~\cite{barrat2011portable}. In summary, a white noise signal is first produced for a window of $N_{\text{seg}}$ timesteps by sampling from a normal distribution with zero mean and unit variance. This signal is then convolved with a kernel derived from the square root of the power spectral density defined in \eqref{eq:qtb_spectrum}, yielding a colored noise with the correct frequency characteristics (see Ref.~\cite{mangaud2019fluctuation} for detailed methodology). Once computed, the random force is applied throughout the segment before a new noise sample is generated for the next interval. The forces are incorporated into the dynamics using a standard thermostatting approach. In our work, we employed the BAOAB integrator \cite{leimkuhler2015molecular} to implement QTB, owing to its robust theoretical foundation, ease of implementation, and demonstrated compatibility with quantum thermal fluctuations.

As mentioned in the main text, ZPEL was monitored via the deviation from zero of Eq.~\ref{eq:qtb_delta_fdt}. No appreciable violation of this condition and/or change in the observables was observed in auxiliary ad-QTB calculations. Those results were therefore not reported.

\section{Removing center-of-mass motion in classical and QTB simulations of molecules}
\label{sec:com_removal}

If one removes center-of-mass translations and rotations when using QTB, it is important to appropriately adjust the computation of the spectra entering into the quantum fluctuation-dissipation relation. To better understand why this is, consider the following. In a scenario where the translational component of the center-of-mass velocity is removed, the velocities of the individual $N$ atoms are modified at each timestep according to the following equation:
\begin{equation}
  \tilde{\mathbf{v}}_i(t) = \mathbf{v}_i(t) - \frac{m_i}{M} \sum_{j=1}^N  \mathbf{v}_j(t)
\end{equation}
where $M = \sum_{i=1}^N m_i$ is the total mass of the system.
The two sides of the FDR computed during a QTB simulation are
\begin{align}
  C_{vF,i}(\omega) \\
  m_i\gamma_i(\omega) C_{vv,i}(\omega)
\end{align}
where $C_{vF,i}$ is the velocity-random force correlation, and $C_{vv,i}$ the velocity autocorrelation for atom $i$. 
If the center-of-mass translational component is removed, then the velocity autocorrelation function $C_{vv,i}(\omega)$ is computed using the modified velocities $\tilde{\mathbf{v}}_i(t)$. In order to be consistent in the computation of the FDR, it is necessary to also augment the random forces in an analogous fashion. This is achieved via the following modification:
\begin{equation}
  \tilde{\mathbf{F}}_i(t) = \mathbf{F}_i(t) - \frac{m_i}{M} \sum_{j=1}^N  \mathbf{F}_j(t)
\end{equation}

If, instead, one wishes to remove global rotations, then the angular velocities of the individual atoms are modified at each timestep according to the following equation:
\begin{equation}
  \tilde{\mathbf{v}}_i(t) = \mathbf{v}_i(t) - \boldsymbol{\omega}_{COM}(t) \times \mathbf{r}_i(t)
\end{equation}
where $\boldsymbol{\omega}_{COM}(t)$ is the angular velocity of the center-of-mass of the system at time $t$. $\boldsymbol{\omega}_{COM}$ can be computed simply as
\begin{equation}
\begin{aligned}
  \boldsymbol{\omega}_{COM}(t) 
    &= \mathbf{I}^{-1}(t) \bigg[ 
       \sum_{i=1}^N m_i \,\mathbf{r}_i(t) \times \mathbf{v}_i(t) \\
    &\quad - \big( \mathbf{r}_{COM}(t) \times \mathbf{P}_{COM}(t) \big)
       \bigg]
\end{aligned}
\end{equation}
where $\mathbf{I}$ ($I_{kl} = \sum_{i=1}^N m_i (|\mathbf{r}_i|^2 \delta_{kl} - r_{i,k} r_{i,l})$ with $k,l=x,y,z$, the Cartesian components of the atomic positions) is the moment of inertia tensor of the system, $\mathbf{r}_{COM}$ is the position of the center-of-mass, and $\mathbf{P}_{COM}$ is the total linear momentum of the system. The term which is subtracted from the sum in the square brackets is the contribution to the angular momentum of the total system arising purely from the translational motion of the center-of-mass. This must be removed in order to isolate the purely rotational part of the total system's motion. 

In the BAOAB scheme, both the B and A step conserve the angular momentum (provided the computed physical forces are accurate enough). The removal of overall angular momentum is thus performed after the update of the velocities in the O step.
In order to be consistent in the computation of the FDR, it is necessary to also augment the random forces in an analogous fashion. 

This is achieved via the following modification:
\begin{equation}
  \tilde{\mathbf{F}}_i(t) = \mathbf{F}_i(t) - m_i(\mathbf{r}_i(t) \times \dot{\boldsymbol{\omega}}_{COM}(t))
\end{equation}
where $\dot{\boldsymbol{\omega}}_{COM}(t)$ is the angular acceleration of the center-of-mass of the system at time $t$. $\dot{\boldsymbol{\omega}}_{COM}(t)$ is computed from the torque, $\tau$ acting on the system as a result of the random forces:
\begin{align}
  \dot{\boldsymbol{\omega}}_{COM}(t) 
    &= \mathbf{I}^{-1}(t) \mathbf{\tau}(t) \\
    &= \mathbf{I}^{-1}(t) \sum_{i=1}^N \bigg[
       \mathbf{r}_i(t) \times \mathbf{F}_i(t) 
       \notag \\[-0.4em]
       &\qquad -\, \mathbf{r}_{COM}(t) \times \mathbf{F}_{COM}(t) 
       \bigg]
\end{align}
where again the term which is subtracted from the sum in the square brackets is the contribution to the torque on the system arising purely from the translational motion of the center-of-mass. This must be removed in order to isolate the torque arising purely from the random forces acting on the system.

This modification of the random force is fully consistent with that of the velocities, setting $\boldsymbol{\omega}_{COM}(t)=\dot{\boldsymbol{\omega}}_{COM}(t) \times dt$.
In practice, since the QTB random forces are generated in segments of length $N_{seg}$, these modifications are performed at the end of each segment, just before computing the correlation functions that enter into the FDR.

\section{Infrared spectra of a selection of additional water clusters}
\label{app:water_clusters_ir_extras}
In this section, we present the infrared spectra of the water clusters \ce{H3O+} , \ce{H(H2O)_4+} (\textit{trans}-zundel), \ce{H(H2O)_4+} (\textit{cis}-zundel), \ce{H(H2O)_4+} (ring), which were omitted from the main text for the sake of brevity. The infrared spectra were computed using the same methodology as described in \secref{sec:theory_and_methods}. All spectra are shown at a temperature of 20K to facilitate comparison with available zero temperature VSCF/VCI and double harmonic theoretical results and low temperature experimental and AIMD data.

\subsection{Hydronium ion/protonated water monomer (\ce{H3O+})}

\begin{figure}[H]
  \centering
  \includegraphics[width=0.5\textwidth]{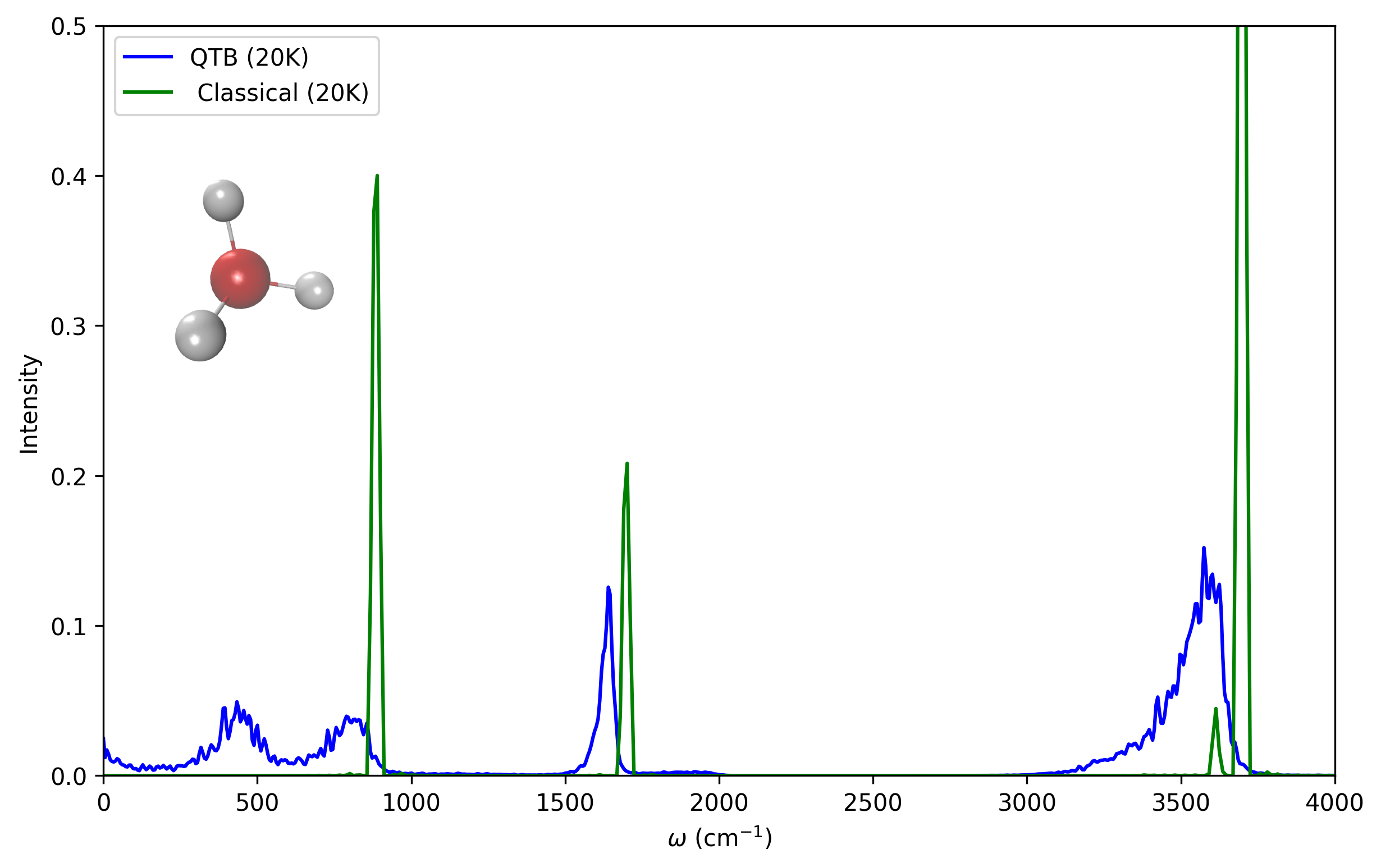}
  \caption{Infrared spectrum of the hydronium ion. Classical results are shown in green while the spectrum obtained via QTB is shown in blue.}
  \label{fig:h3o_ir}
\end{figure}

The infrared spectrum of the hydronium ion is shown in \figref{fig:h3o_ir}. We compare this with the work of Kulig \etal{} \cite{kulig2014both} who report both experimental and theoretical infrared spectra of \ce{H3O+}. Their (classical) theoretical spectra were computed at $T=50$K using AIMD with the BLYP functional, a double-$\zeta$ valence polarization (DZVP) basis set and Goedecker-Teter-Hutter pseudopotential.

Looking first at the far-infrared region, we note the two vibration-inversion bands (($\nu_2 (1^{-} \leftarrow 0^{+})$ and $\nu_2(1^{+} \leftarrow 0^{+}))$ are clearly visible in the QTB results in Fig.~\ref{fig:h3o_ir} at around 820 \cminv and 450 \cminv respectively. The experimental results reported by Kulig \etal{} and previous experimental and theoretical results (see Ref.~\cite{liu1986calculated}) place the centers of these bands at around 925 \cminv{} and 525 \cminv{}. On the other hand, the AIMD results of Kulig \etal{} fail to predict two distinct bands in this region, instead showing a single band at 813 \cminv{} (see Figure S3 in the supporting information of Ref.\cite{kulig2014both}). This is likely due to their neglect of nuclear quantum effects. Indeed, given that the $\nu_2 (1^{-} \leftarrow 0^{+})$ and $\nu_2(1^{+} \leftarrow 0^{+}))$ modes arise from umbrella inversion motion of the hydronium ion, it is expected that these modes will be significantly affected by nuclear quantum effects which facilitate tunnelling even at low temperatures. Contrariwise, for these modes to show up in classical simulations, it is necessary to have a sufficiently high temperature to overcome the energy barrier associated with the inversion motion. Indeed, these IR modes do in fact appear in the classical MD spectrum when the temperature is raised to 300 K.  

The next prominent band appearing in the spectrum of the hydronium ion is predicted by experiment to be located at around 1622 \cminv{}. Our QTB results predict a band at $\sim$1640 \cminv{} which is in good agreement with this experimental value. Contrariwise, we find our classical results predict the center of this band to be located at closer to $\sim$1700 \cminv{} indicating that NQEs act to noticeably red-shift this band. A similar observation is made for the band located at $\sim$3510 \cminv{} in the experimental results  (corresponding to stretching modes). Our QTB results predict this peak to be located at $\sim$3580 \cminv{}, although this peak is very broad and extends from $\sim$3000 \cminv{} to $\sim$3800 \cminv{} which we attribute to the broadening effects of the thermostat. The classical results, on the other hand, predict a main peak in this region at $\sim$3700 \cminv{} which again indicates the importance of incorporating NQEs to accurately predict the position of this band.

\subsection{\ce{H(H2O)_4+} in the \textit{cis}-zundel and \textit{trans}-zundel conformations}

\begin{figure}[H]
  \centering
  \includegraphics[width=0.5\textwidth]{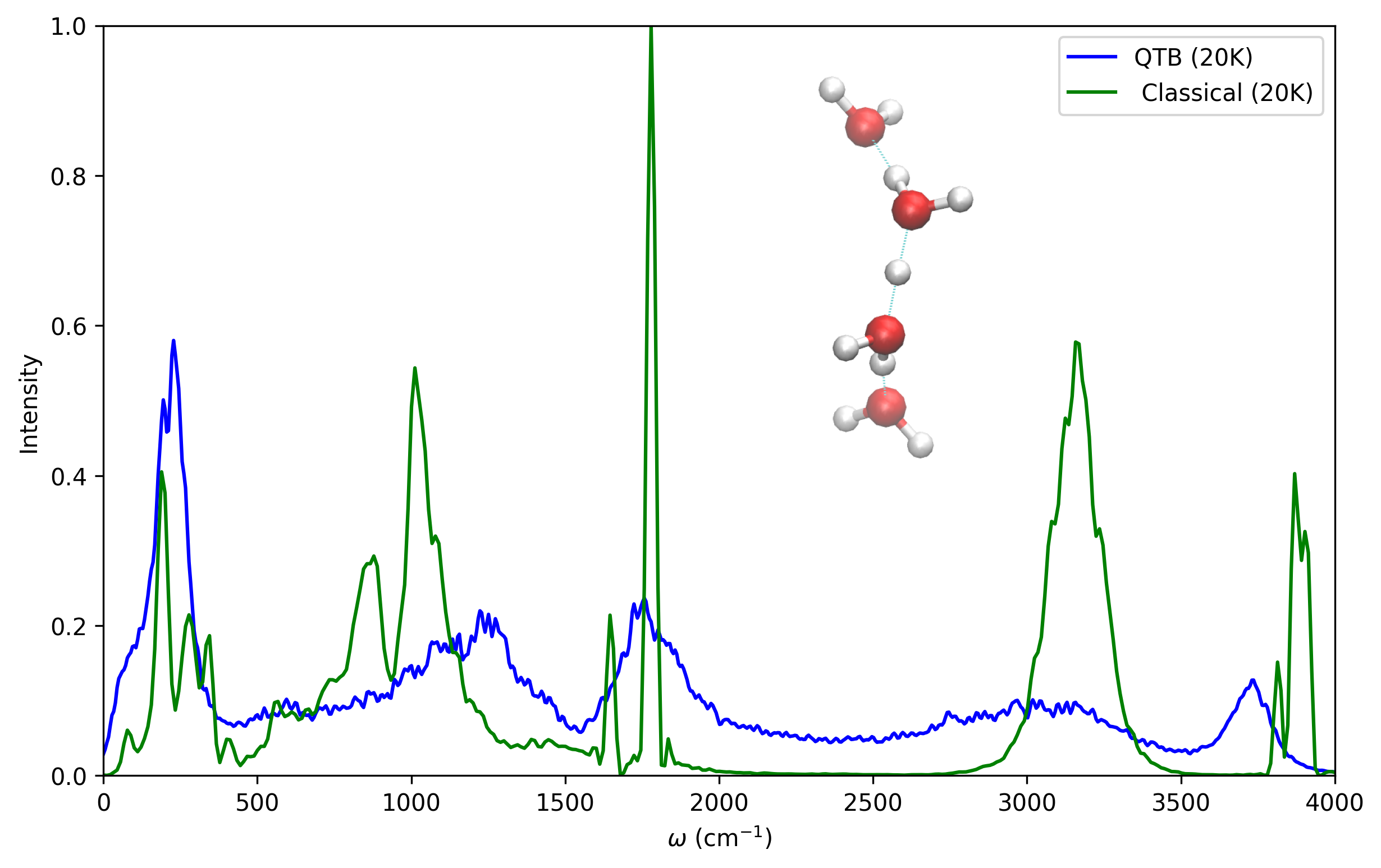}
\caption{Infrared spectrum of the \textit{cis}-Zundel conformation of \ce{H(H2O)_4+} at a temperature of 20K.Classical results are shown in green while the spectrum obtained via QTB is shown in blue.}
\label{fig:h9o4_cis_zundel_ir}
\end{figure}

\begin{figure}[H]
  \centering
  \includegraphics[width=0.5\textwidth]{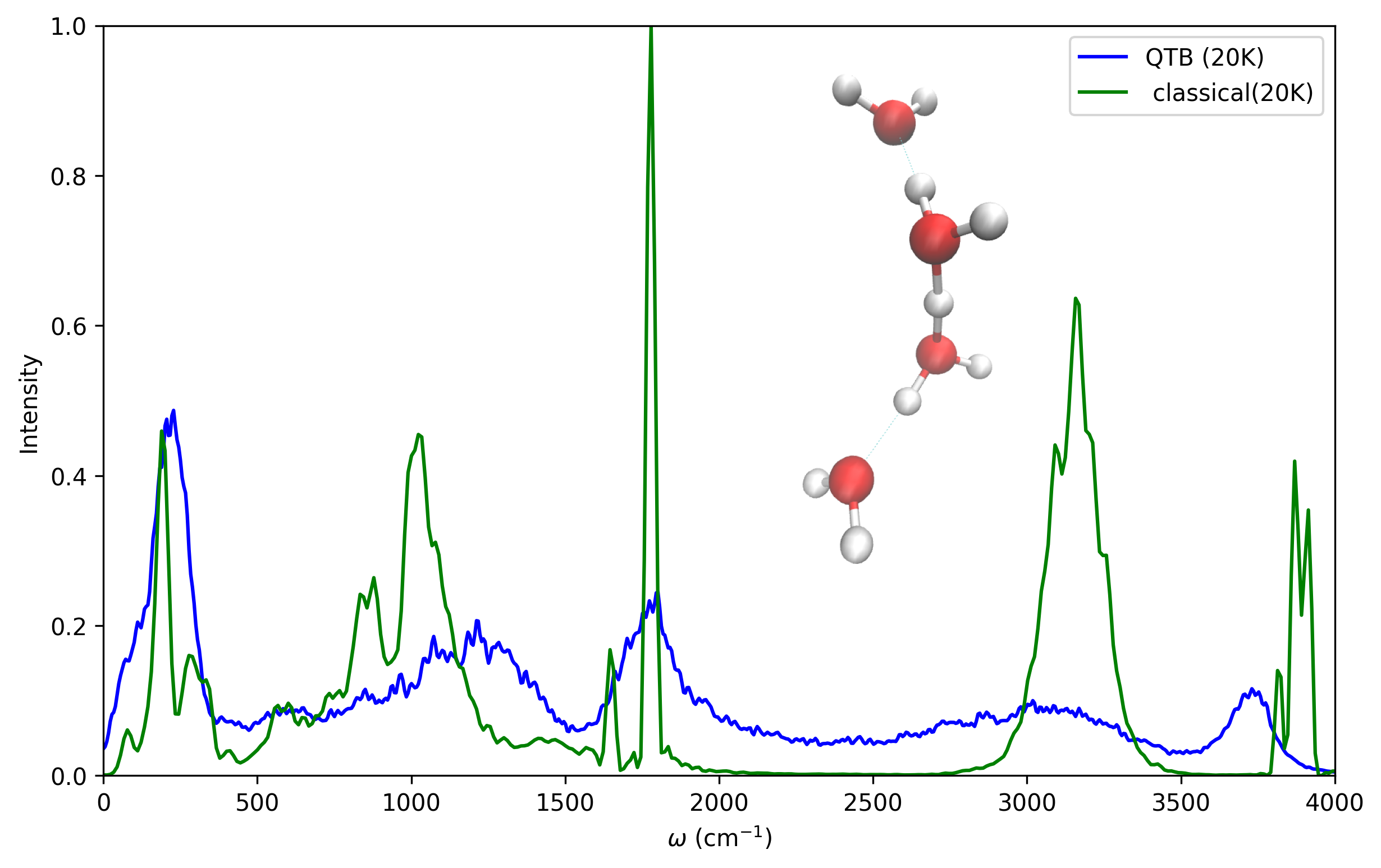}
\caption{Infrared spectrum of the \textit{trans}-Zundel conformation of \ce{H(H2O)_4+} at a temperature of 20K. Classical results are shown in green while the spectrum obtained via QTB is shown in blue.}
\label{fig:h9o4_trans_zundel_ir}
\end{figure}

IR spectra of the \textit{cis}- and \textit{trans}-Zundel conformations of \ce{H(H2O)_4+} are not frequently reported in the literature. This is ostensibly due to the fact that these conformations of \ce{H(H2O)_4+} are less energetically favourable than the Eigen conformation discussed in the main text (see Sec.~\ref{sec:eigen_spectra}). However, the investigation of these conformations (along with the ring conformation discussed next) is important in disentangling the contributions of these conformers to the overall spectrum of \ce{H(H2O)_4+}. As noted by Kulig \etal{}, many of the spectral features in the low frequency region of \ce{H(H2O)_4+} are not possible to explain without considering the contributions of these conformers. We therefore present the calculated spectra of the \textit{cis}- and \textit{trans}-Zundel conformations of \ce{H(H2O)_4+} in \figref{fig:h9o4_cis_zundel_ir} and \figref{fig:h9o4_trans_zundel_ir} respectively.

The spectra of both the \textit{cis}-Zundel and \textit{trans}-Zundel conformations unsurprisingly exhibit many of the same features found in the spectrum of the \ce{H5O2+} (Zundel) ion discussed in \secref{sec:zundel_ir}. Indeed, as noted by Yu \etal{} in Ref.~\cite{yu2017high}, the dominant band in the spectra of the Zundel isomers is located at 1000 \cminv{} and is due to stretching motion of the shared proton within the isomers. We find that our classical results also predict a band located at $\sim$1000 \cminv{}. However, our QTB spectra show this peak to be noticeably blue-shifted to $\sim$1250 \cminv{}. This is consistent with our previous observations for the Zundel ion (see \secref{sec:zundel_ir}).

\subsection{\ce{H(H2O)_4+} (ring conformation)}

\begin{figure}[H]
  \centering
  \includegraphics[width=0.5\textwidth]{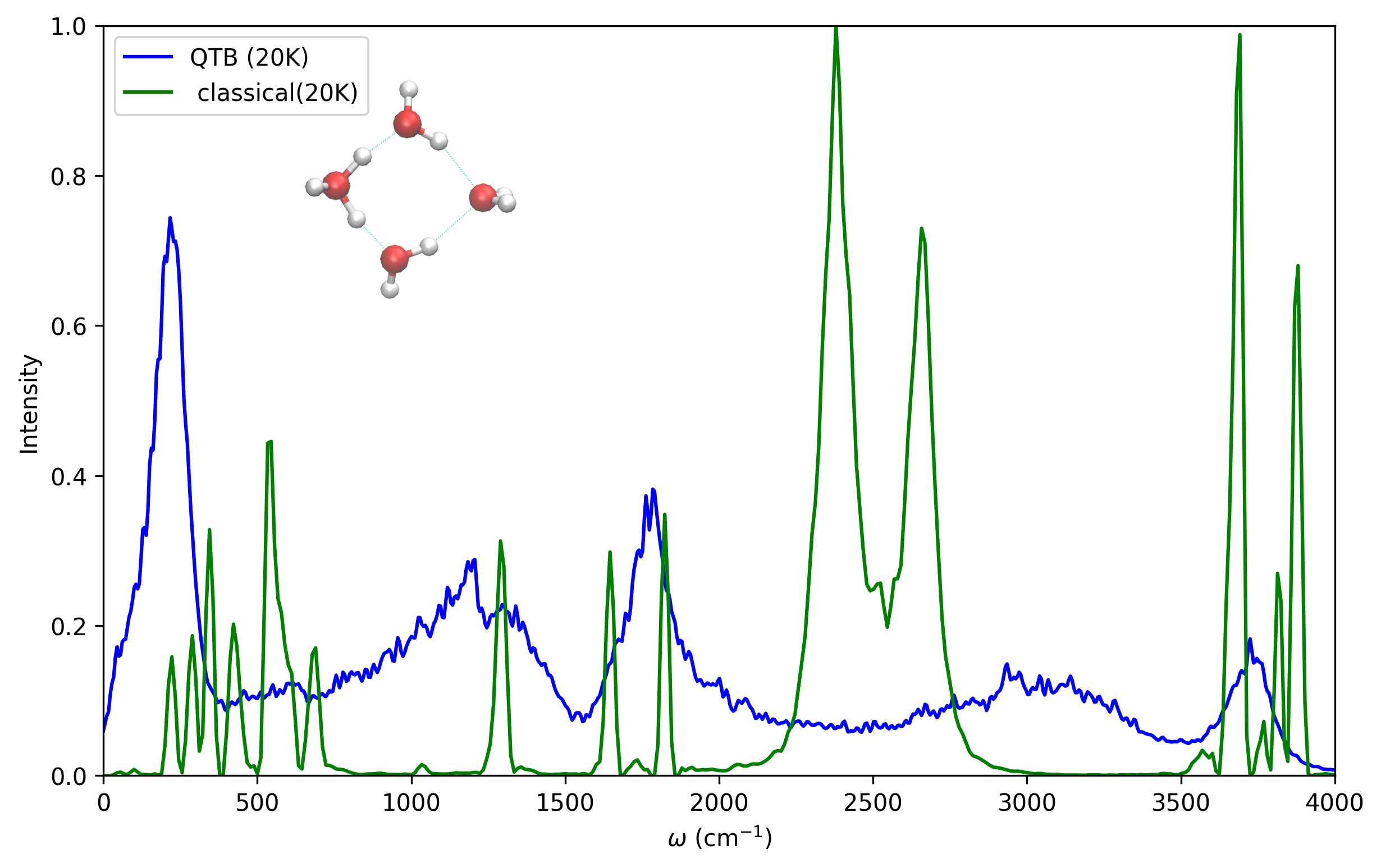}
\caption{Infrared spectrum of the ring conformation of \ce{H(H2O)_4+} at a temperature of 20K.Classical results are shown in green while the spectrum obtained via QTB is shown in blue.}
\label{fig:h9o4_ring_ir}
\end{figure}

We present our calculated IR spectra for the ring conformation of \ce{H(H2O)_4+} in \figref{fig:h9o4_ring_ir}.
As noted by Yu \etal{}, the ring conformation can be thought of as a water molecule hydrogen bonded to an $\ce{H7O3+}$ cluster, which results in the ring-like structure. Consequently, one expects to observe many of the features of these two species within the spectrum of the ring isomer. This is indeed the case, and the spectrum of the ring conformation of \ce{H(H2O)_4+} exhibits bands at approximately 1450, 1750, 2000, and 3660 \cminv{} which coincide with those of \ce{H7O3+}. These bands have been attributed to a water bending mode, hydronium bending mode, hydronium stretching mode, and the OH stretching mode of hydronium respectively. In addition to these bands, the VSCF/VCI calculations performed by these authors also predict the presence of several less intense features in the frequency ranges bridging the aforementioned bands. Looking at the QTB spectra (blue curve in Fig.~\ref{fig:h9o4_ring_ir}) it would appear that these features are responsible for the broad bands present in the resulting spectrum, whose peaks are centered at around 1150 \cminv{}, 1750 \cminv{}, 3000 \cminv{} and 3750 \cminv{}. Due to the spectral broadening of the QTB thermostat, these features tend to coalesce into these broad bands.

\section{Timings}
\label{sec:watercluster_timings}
As an illustration of the computational efficiency of the QTB and adQTB methods, we present in Table.\ref{tab:timings} the average timings per MD step for each of the 8 different water clusters considered, and each of the three different methods (classical MD, QTB, and adQTB) as well as an estimation of the total MD trajectory length that can be attained on standard HPC hardware in 24 hours of wall time. The data for adQTB is provided for completeness, even though - as previously explained - the use of this method was not necessary for the spectra presented in the paper. The simulations were performed on a single core of an Intel(R) Xeon(R) Platinum 8360Y CPU @ 2.40GHz. The timings were averaged over the entire trajectory length of 1ns, and the number of steps taken was $10^7$. The scaling with system size is approximately linear, with QTB and adQTB generally taking around the same wall time per MD step as classical MD. As a consequence, QTB constitutes a viable method for the inclusion of NQEs in larger molecular systems, facilitating access to the long simulation times necessary to accumulate sufficient statistics for accurate resolution of IR spectra. 


\begin{table}[H]
    \centering
    \caption{Average timings per MD step for each of the eight different water clusters considered, using three different methods (Classical MD, QTB, and adQTB). The last column indicates the duration (in nanoseconds) of the MD trajectory that can be simulated per day on a single core of an Intel(R) Xeon(R) Platinum 8360Y CPU @ 2.40GHz.}
    \label{tab:timings}
    \begin{tabular}{|l|l|c|c|}
        \hline
        \textbf{Molecule Name} & \textbf{Method} & \textbf{\shortstack{Avg Time\\ per Step (s)}} & \textbf{ns per Day} \\ \hline
        \ce{H2O} & Classical & 0.000116 & 86.207 \\ \hline
        \ce{H2O} & QTB & 0.000115 & 86.957 \\ \hline
        \ce{H2O} & adQTB & 0.000107 & 93.458 \\ \hline
        \ce{H3O+} & Classical & 0.000142 & 70.423 \\ \hline
        \ce{H3O+} & QTB & 0.000146 & 68.493 \\ \hline
        \ce{H3O+} & adQTB & 0.000143 & 69.930 \\ \hline
        \ce{H(H2O)2+} & Classical & 0.000317 & 31.549 \\ \hline
        \ce{H(H2O)2+} & QTB & 0.000322 & 31.055 \\ \hline
        \ce{H(H2O)2+} & adQTB & 0.000316 & 31.646 \\ \hline
        \ce{H(H2O)3+} & Classical & 0.000546 & 18.315 \\ \hline
        \ce{H(H2O)3+} & QTB & 0.000568 & 17.609 \\ \hline
        \ce{H(H2O)3+} & adQTB & 0.000549 & 18.221 \\ \hline
        \ce{H(H2O)4+} (Eigen) & Classical & 0.000962 & 10.398 \\ \hline
        \ce{H(H2O)4+} (Eigen) & QTB & 0.000776 & 12.890 \\ \hline
        \ce{H(H2O)4+} (Eigen) & adQTB & 0.000852 & 11.737 \\ \hline
        \ce{H(H2O)4+} (\textit{trans}-zundel) & Classical & 0.000807 & 12.384 \\ \hline
        \ce{H(H2O)4+} (\textit{trans}-zundel) & QTB & 0.000848 & 11.789 \\ \hline
        \ce{H(H2O)4+} (\textit{trans}-zundel) & adQTB & 0.000850 & 11.765 \\ \hline
        \ce{H(H2O)4+} (\textit{cis}-zundel) & Classical & 0.002526 & 3.956 \\ \hline
        \ce{H(H2O)4+} (\textit{cis}-zundel) & QTB & 0.002931 & 3.409 \\ \hline
        \ce{H(H2O)4+} (\textit{cis}-zundel) & adQTB & 0.002941 & 3.400 \\ \hline
        \ce{H(H2O)4+} (ring) & Classical & 0.000587 & 17.012 \\ \hline
        \ce{H(H2O)4+} (ring) & QTB & 0.000852 & 11.737 \\ \hline
        \ce{H(H2O)4+} (ring) & adQTB & 0.002258 & 4.428 \\ \hline
    \end{tabular}
\end{table}

\end{document}